\documentstyle[aaspptwo,psfig]{article}

\def\etal{{\it et al. }}

\begin{document}

\title{Globular Cluster Luminosity Functions and the Hubble Constant
from WFPC2 Imaging: The Giant Elliptical NGC 4365\altaffilmark{1}}

\author{Duncan A. Forbes}
\affil{Lick Observatory, University of California, Santa Cruz, CA 95064}
\affil{Electronic mail: forbes@lick.ucsc.edu}

\altaffiltext{1}{Based on observations with the NASA/ESA {\it Hubble
Space Telescope}, obtained at the Space Telescope Science Institute,
which is operated by AURA, Inc., under NASA contract NAS 5--26555}

\begin{abstract}

The turnover, or peak, magnitude in a galaxy's globular cluster luminosity
function (GCLF) may provide a standard candle for an independent distance
estimator. Here we examine the GCLF of the giant elliptical NGC 4365 using
photometry of $\sim$ 350 globular clusters from the {\it Hubble Space
Telescope's} Wide Field and Planetary Camera 2 (WFPC2). The WFPC2 data have
several advantages over equivalent ground--based imaging. The membership of NGC
4365 in the Virgo cluster has been the subject of recent debate. We have fit a
Gaussian and $t_5$ profile to the luminosity function and find that it can be
well represented by a turnover magnitude of $m_V^0$ = 24.2 $\pm$ 0.3 and a
dispersion $\sigma$ = 1.28 $\pm$ 0.15. After applying a small metallicity
correction to the `universal' globular cluster turnover magnitude, we derive a
distance modulus of (m -- M) = 31.6 $\pm$ 0.3 which is in reasonable agreement
with that from surface brightness fluctuation measurements. This result places
NGC 4365 about 6 Mpc beyond the Virgo cluster core. For a V$_{CMB}$ = 1592$\pm$
24 km s$^{-1}$ the Hubble constant is H$_{\circ}$ = 72$_{-12}^{+10}$ km
s$^{-1}$ Mpc$^{-1}$. We also describe our method for estimating a local
specific frequency for the GC system within the central 5 h$^{-1}$ kpc which
has fewer uncertain corrections than a total estimate. The resulting value of
6.4 $\pm$ 1.5 indicates that NGC 4365 has a GC richness similar to other early
type galaxies.  

\end{abstract}


\section{Introduction}

The measurement and interpretation of extragalactic distances are problematic
and often the subject of dispute. In a review of techniques for measuring
distances, Jacoby \etal (1992) describe five methods that can be applied to
elliptical galaxies, namely planetary nebula luminosity functions (PNLF),
novae, surface brightness fluctuations (SBF), the D$_n$ -- $\sigma$ relation
and globular cluster luminosity functions (GCLF). In the case of the GCLF
method, the `standard candle' is the magnitude of the turnover, or peak, in the
luminosity function. Although there is no generally accepted theoretical basis,
all well--studied globular cluster (GC) systems reveal a similar GCLF shape
(often approximated by a Gaussian) with a turnover magnitude of M$_V^0$ $\sim$
--7.5.  Measurement of distances, and hence the Hubble constant H$_{\circ}$,
with this method, rely on the assumption that the turnover is the same for all
galaxies. Recently there have been claims that the turnover magnitude is not
quite constant, but has a slight dependency ($\sim$ 0.2 mag) on GC metallicity
(Ashman, Conti \& Zepf 1995) or the dispersion in the GCLF, which in turn may
reflect a galaxy Hubble type dependence (Secker \& Harris 1993).  

There have been two recent developments of direct relevance to GCLF--determined
distances. The first is due to the much improved Strehl ratio (image
concentration) of the second Wide Field and Planetary Camera (WFPC2) on the
{\it Hubble Space Telescope}. Now even relatively short exposures of
ellipticals with WFPC2 can contain hundreds of GCs, several magnitudes fainter
than typical ground--based observations. To date, published distance
measurements based on WFPC2 studies of the GCLF have been carried out by Baum
\etal (1996) on NGC 4881 in Coma and by Whitmore \etal (1995) on M87 in Virgo.
These studies quote values of H$_{\circ}$ = 67 and 78 km s$^{-1}$ Mpc$^{-1}$
respectively. The second development is a new calibration of Galactic GC
distances based on revised RR Lyrae luminosities by Sandage \& Tammann (1995)
which gives an absolute turnover magnitude of M$_V^0$ = --7.60 $\pm$ 0.11 for
our Galaxy. Combined with the halo GCs in M31, they derive M$_V^0$ = --7.62
with an internal error of $\pm$ 0.08 and external error of $\pm$ 0.2 mags.  

Sandage \& Tammann (1995) went on to re--analyze ground--based GCLFs for five
Virgo ellipticals. Although they derive a distance modulus similar to that of
Secker \& Harris (1993) for Virgo, they disagree on a couple of issues. In
particular, Sandage \& Tammann question the dependence of M$_V^0$ on the GCLF
dispersion claiming that it is in part an artifact of the fitting procedure.
They also disagree on the cluster membership of one galaxy -- NGC 4365. Secker
\& Harris claim that it lies slightly more distant than the Virgo cluster in
the W cloud, as supported by the surface brightness fluctuation (SBF)
measurements of Tonry, Ajhar \& Luppino (1990).  Sandage \& Tammann (1995), on
the other hand, suggest that the high metallicity of NGC 4365 makes the SBF
method unreliable.  Both of these GCLF studies used data from Harris \etal
(1991) and they both derived a $m_V^0$ for NGC 4365 to be about 0.8 mags
fainter than typical Virgo ellipticals. However the uncertainty on this value
is much larger than for the other Virgo ellipticals in the Harris \etal (1991)
data set.  

A WFPC2 study of the GCLF for NGC 4365 would have several advantages to help
resolve the issue of Virgo membership and further test the hypothesis of a
universal GCLF. As well as providing an independent data set, the benefits
include very low background contamination, no serious blending effects,
accurate photometry and the ability to probe to faint magnitude levels.  The
galaxy itself is also of interest as it has a relatively high Mg$_2$ index
(Davies \etal 1987), contains a kinematically--distinct core  which is
detectable as a disk--like structure in both the kinematics (Surma 1992) and
photometry (Forbes 1994), and has a notably blue nucleus (Carollo \etal 1996).  

Forbes \etal (1996) presented WFPC2 data on the GCs in 14 ellipticals with
kinematically--distinct cores (KDC). In that paper we discussed the colors,
radial and azimuthal distribution of the GCs. Analysis of the GCLFs were not
attempted. Here we analyze the GCLF of NGC 4365, the richest GC system in the
Forbes \etal study.  After determining a completeness function and quantifying
the photometric errors, we use the maximum likelihood method of Secker \&
Harris (1993) to determine the turnover magnitude and dispersion of the GCLF.
Within the assumptions of the GCLF--distance method, this leads to an
independent estimate of the Hubble constant H$_{\circ}$.  

\section{Observations and Data Reduction}

Details of the WFPC2 data for the NGC 4365 GCs are presented, along with 13
other KDC ellipticals, in Forbes \etal (1996). Briefly, two 500s F555W images
were combined, as were two 230s F814W images. Using DAOPHOT (Stetson 1987), GCs
were detected in the F555W image only, their magnitudes measured and then the
corresponding F814W magnitudes were determined. The magnitudes have been
converted into the standard Johnson--Cousins V, I system and corrected for
Galactic extinction. We chose a fairly conservative detection criteria based on
flux threshold, shape, sharpness and size. Additionally, we checked the
positions of GCs against a list of known hot pixels. After these criteria have
been applied, we are confident that the contamination from cosmic rays, hot
pixels, foreground stars and background galaxies is small (less than a few
percent) in our object list.  Forbes \etal (1996) made one additional cut,
namely $\pm$3$\sigma$ about the mean color of V--I = 1.10. Here we have chosen
to use only the V band data (which is of higher signal-to-noise than the I band
data) and not apply any selection based on color.  

The detection flux threshold in the PC image was set lower than the WFC images,
to compensate for the different point source sensitivity (i.e. $\sim$0.3 mags;
Burrows \etal 1993). In Figure 1 we show the fraction of actual GCs detected as
a function of GC magnitude, normalized at V = 25.  This figure shows that the
resulting detection fractions are similar between all four CCDs.  

\section{Modeling}

\subsection{Completeness Function}

Forbes \etal (1996) carried out simulations to quantify the ability of DAOPHOT
to detect GCs as a function of magnitude. A typical WFC image was chosen for
the simulation. The resulting completeness function showed that all GCs
brighter than V $\sim$ 24 were detected, and the completeness dropped off
rapidly to V $\sim$ 25. As it is crucial for GCLF studies to have a
well--determined completeness function we have decided to redo the simulation
using the WF3 image of NGC 4365 to ensure that we have the same photon and read
noise characteristics as the data. We note that there is no evidence for a
significant variation between CCDs. We have simulated GCs using the {\it
addstar} task and then used DAOPHOT with the same detection criteria as for the
actual GCs. In particular, we have excluded all objects with FWHM sizes greater
than three pixels. For these simulations $\le$ 1\% of the objects are excluded
by this criterion. As with the real data, we have not attempted to reject any
objects based on color.  

The completeness function resulting from simulations of 900 artifical GCs is
shown in Figure 2a.  This function is similar to that given in Forbes \etal
(1996).  We derive a 50\% completeness level at V = 24.7. For the subsequent
analysis the completeness function is set to zero for magnitudes fainter than
this to avoid incompleteness corrections larger than a factor of two.  

\subsection{Photometric Errors}

In addition to the completeness function we need to quantify the photometric,
or measurement, error from DAOPHOT. Photometric errors can cause a shift of the
GCLF peak to brighter magnitudes, as the fainter GCs, with relatively large
errors, move into brighter magnitude bins. This `bin jumping' effect is
described in detail by Secker \& Harris (1993) and taken into account in their
maximum likelihood technique. Here we have fit the DAOPHOT determined errors
with an exponential of the form:\\

p.e. = exp~[a~(V~--~b)]\\

\noindent The photometric error and the fit as a function of V magnitude are
shown in Figure 2b. Reassuringly these errors are similar to those found by
comparing the input and measured magnitudes of the simulated GCs. A typical
photometric error is $\pm$ 0.1 mag at V = 24.  

\subsection{Background Contamination}

One of the advantages of using WFPC2 data for GCLF studies is the ability to
exclude most foreground stars and background galaxies based on angular size.
This means that the contamination from such sources is very low. Nevertheless
we estimated the background contamination on a similar exposure time WFPC2
image from the Medium Deep Survey (Forbes \etal 1994). Again we used DAOPHOT to
detect objects with the same detection parameters as before, including the same
size criteria as for the GCs. No color selection was used.  We estimate a
background contamination of seven objects, brighter than V = 24.7, in the WFPC2
field-of-view.  

\subsection{Maximum Likelihood Technique}

In this study we use the maximum likelihood technique of Secker \& Harris
(1993) to accurately determine the GCLF peak magnitude and dispersion. Their
technique is designed to take proper account of detection incompleteness at
faint magnitudes, photometric error and background contamination.  It
calculates the convolution product of the photometric error and the intrinsic
GCLF, weighted by the completeness function. This is then compared to the raw
data set, and after allowing for the background contamination, gives the most
likely parameters of the intrinsic GCLF.  As well as the commonly used Gaussian
profile, we fit a $t_5$ distribution which is less susceptible to variations at
the extremes of the luminosity function. The $t_5$ distribution function has
the form:\\

N$(m) = A (1 + (m - m^0 )^2 / 5
\sigma_t^2)^{-3}$\\

\noindent
Where A is a scaling constant and $\sigma_t$ is the GCLF dispersion, which is
related to the dispersion of a Gaussian by $\sigma_t \sim 0.78 \sigma_G$ (see
Secker 1992 for details of the $t_5$ function).  

\section{Results and Discussion}

After applying the maximum likelihood code to our sample of 346 GCs with V $<$
24.7, we find the best estimate and uncertainty for a Gaussian profile fit to
the GCLF to be $m_V^0$ = 24.17 (+0.3,--0.3), $\sigma$ = 1.36 (+0.14,--0.15).
For a $t_5$ profile fit, we find $m_V^0$ = 24.00 (+0.3,--0.2), $\sigma$ = 1.17
(+0.15,--0.13). These errors represent the collapsed one--dimensional
confidence limits for one standard deviation.  The probability contours output
from the maximum likelihood code for the Gaussian fit, over a range of 0.5--3
standard deviations, are shown in Figure 3. The contours are skewed towards a
larger dispersion and fainter magnitudes, giving rise to the asymmetric errors
quoted above. A similar effect can be seen in the ground--based data of NGC
4365 by Secker \& Harris (1993). Although the quoted errors represent the
internal error of the fitting procedure, they dominate over any contribution
from photometric errors. As a test, we increased the photometric errors by 20\%
(which represents the extreme range of photometric errors from DAOPHOT) and
refit the data. This gave a turnover magnitude and dispersion different by $<$
3\%.  In Figure 4 we show a binned GCLF and our best--fit Gaussian superposed.
Note that the fitting procedure does not use binned data but rather treats each
data point individually.  

Secker \& Harris (1993) have shown that the GCLF parameters will be
systematically biased towards brighter magnitudes and smaller dispersions if
the limiting magnitude is close to or brighter than the true turnover
magnitude. Our limiting magnitude has been set at the 50\% completeness level,
i.e. V = 24.7. Using their figure 6, the true turnover magnitude for a $t_5$
distribution is $\sim$ 0.1 mag fainter and the dispersion 0.05 mag larger.  The
results of the two fitting methods, after applying this bias correction to both
results, are listed in Table 1.  Averaging the results from the two fitting
methods, gives $m_V^0$ = 24.2 $\pm$ 0.3 and $\sigma$ = 1.28 $\pm$ 0.15, which
is a reasonable representation of the turnover and dispersion of our V band
data for the GCLF of NGC 4365.  We also list the results of Secker \& Harris
for the B band GCLF.  If we assume that the NGC 4365 GCs have B--V $\sim$ 0.9,
then their turnover is fainter by $\sim$ 0.2 mags.  We find a slightly smaller
GCLF dispersion. The quoted errors are similar between the two studies.  

An additional small correction may be required if we wish to compare the
magnitudes of the NGC 4365 GCs with the `universal value' (i.e. the mean of the
Milky Way and M31 systems from Sandage \& Tammann 1995).  As mentioned in the
introduction, Secker \& Harris (1993) and also Fleming \etal (1995) suggest
that the GCLF turnover depends on Hubble type. Compared to nearby spirals, the
more luminous ellipticals have a fainter turnover. On a more quantitative
basis, Ashman \etal (1995) have suggested that GC metallicity is the second
parameter, with more metal rich GCs having a fainter turnover. These ideas can
be connected via the GC metallicity--galaxy luminosity relation (e.g. Brodie \&
Huchra 1991, Forbes \etal 1996) in which more luminous galaxies (e.g. giant
ellipticals) have relatively metal rich GC systems.  Ashman \etal showed that
if these metallicity--based corrections were applied to the turnover GC
absolute magnitude, then the systematic offset between GCLF distance estimates
and other methods (see Jacoby \etal 1992) could be largely eliminated.  

In the absence of spectroscopic measures, the mean metallicity of the GC system
in NGC 4365 can be estimated crudely from the V--I colors of the GCs. Using the
GC sample described in section 2, we calculate a mean metallicity, assuming
[Fe/H] = 5.051 (V--I) -- 6.096 (Couture \etal 1990), of [Fe/H] = --0.6. The
mean metallicity of the Milky Way and M31 GCs, using the same relative
weighting as Sandage \& Tammann (1995) is [Fe/H] = --1.4. Applying Table 3 of
Ashman \etal gives $\Delta$M$_B^0$ = 0.37 and $\Delta$M$_V^0$ = 0.23 for a
metallicity difference of 0.8 dex.  (A metallicity difference of 0.7--0.9 dex
would correspond to roughly $\Delta$M$_B^0$ = 0.32--0.45 and $\Delta$M$_V^0$ =
0.18--26.) These corrections make the combined Milky Way and M31 peaks of
M$_B^0$ = --6.93 $\pm$ 0.08 and M$_V^0$ = --7.62 $\pm$ 0.08 (Sandage \& Tammann
1995) fainter by 0.37 and 0.22 respectively. Combining these M$^0$ values with
$m^0$ from Table 1 gives the distance modulus for both the Gaussian and $t_5$
fits.  The fits and the averages (with errors added in quadrature and divided
by $\sqrt{N}$) are listed in Table 2. From our data the $t_5$ and Gaussian fits
give $(m - M)$ = 31.6 $\pm$ 0.3.  

An independent estimate of the distance modulus comes from surface brightness
fluctuation (SBF) measurements (Tonry, Ajhar \& Luppino 1990). Ajhar \etal
(1994) quote an SBF distance modulus of 31.74 $\pm$ 0.16 for NGC 4365, based on
the latest calibration of Tonry (1991), which is given in Table 2. The two GCLF
distance determinations are in good agreement with that from the SBF method.  

We also list in Table 2, recent determinations for the distance modulus to the
Virgo galaxies NGC 4472 (M49), NGC 4486 (M87) and NGC 4649 using the GCLF
method. Again the distance modulus is calculated using either M$_B^0$ = --6.93
$\pm$ 0.08 or M$_V^0$ = --7.62 $\pm$ 0.08 and a metallicity correction from
Ashman \etal (1995), with GC mean metallicities as compiled by Perelmuter
(1995).  These three galaxies give a GCLF distance modulus to Virgo of about
$(m - M)$ = 31.25 $\pm$ 0.15 (internal error only), which can be compared to
the weighted mean from 6 different methods (i.e. novae, SN Ia, Tully--Fisher,
PNLF, SBF and D$_n$--$\sigma$) of $(m - M)$ = 30.97 $\pm$ 0.18.  

To summarize, the GCLF method indicates a similar distance to the Virgo cluster
as other distance methods, and the GCLF and SBF distances to NGC 4365 are in
good agreement. However, the distance modulus for NGC 4365 is 0.5--0.75
magnitudes fainter than that for the Virgo core. This would suggest it is
25--40\% more distant than the Virgo core.  Using a representative GCLF
distance modulus of $(m - M)$ = 31.74 $\pm$ 0.3 gives a distance of 22.28
(+3.31, --2.87) Mpc.  

It is of course interesting to take the distance calculation one step further
and estimate the Hubble constant from this one galaxy.  The velocity of NGC
4365 with respect to the cosmic microwave background is V$_{CMB}$ = 1592 km
s$^{-1}$ (Faber \etal 1989). We assume an error on this value to be the
fractional error from the radial velocity measurement by Huchra \etal (1983),
i.e. $\pm$ 24 km s$^{-1}$. Dividing this velocity by 22.28 Mpc gives a Hubble
constant of 72 (+10, --12) km s$^{-1}$ Mpc$^{-1}$.  The velocity from the
D$_n$--$\sigma$ relation is similar, i.e.  1509 $\pm$ 250 km s$^{-1}$ after
making a $\sim$3\% correction for the Malmquist bias (Faber \etal 1989).  This
would give a Hubble constant of 68 (+21, --20) km s$^{-1}$ Mpc$^{-1}$.  

Another measure of interest is the specific frequency ($S$) of the GC system,
which gives an indication of the relative richness of the GC system, and is
defined by:\\

$S = N 10^{0.4(M_V + 15)}$

\noindent
Where N is usually the total number of GCs and M$_V$ the total galaxy
magnitude. Estimates of $S$ for galaxies beyond the local group require two,
sometimes large and uncertain, corrections for the number of faint GCs that
weren't detected and the limited areal coverage.  Starting with the first
correction, by integrating under our profile fit to the GCLF we can make a
fairly accurate estimate of the total number of GCs within the WFPC2
field-of-view. The $t_5$ and Gaussian fits give a total of 522 and 554 GCs
respectively, over all magnitudes. Taking the average of these we get 538 GCs. 
We find that 84.7\% of the NGC 4365 GCs lie within a 180$^{\circ}$ hemisphere
of radius 100$^{''}$ (5~h$^{-1}$ kpc).  The total number of GCs within a
100$^{''}$ radius circle is twice this amount or 911 (with an estimated error
of $\pm$12\%).  Knowing the integrated galaxy absolute magnitude within this
radius will give us a `local' $S$ value. From the surface photometry of
Goudfrooij \etal (1994), we calculate a magnitude of V = 11.36 $\pm$ 0.1, and
using $(m - M)$ = 31.74 $\pm$ 0.3, gives a localized specific frequency of $S$
= 6.4 $\pm$ 2.7.  The second correction, calculating the total number in the GC
system, is much more uncertain.  This can be estimated by integrating the
density profile found by Forbes \etal (1996) out to large radii, with the
boundary condition that at r = 100$^{''}$, the number of GCs is 911. This gives
a total for the GC system of N = 2511 $\pm$ 1000, which can be compared to N =
3500 $\pm$ 1200, from ground--based imaging, estimated by Harris (1991).  Using
a total V = 9.65 $\pm$ 0.1 (Faber \etal 1989) and the same distance modulus as
above, gives $M_V$ = --22.09 $\pm$ 0.32 and $S$ = 3.7 $\pm$ 2.4. Our estimated
$S$ values, 6.4 and 3.7, are similar to the average value of 5.1 for 34 E+S0
galaxies (van den Bergh 1995). Harris (1991) quoted 7.7 $\pm$ 2.7 for NGC 4365.
Part of the difference is due to our lower number of GCs and also because
Harris, assumed that NGC 4365 was in Virgo with a distance modulus of 31.3. For
N = 3500 $\pm$ 1200 and our a distance modulus of 31.74 $\pm$ 0.3, $S$ = 5.1
$\pm$ 3.3.  

\section{Conclusions}

We have used the {\it HST} WFPC2 data of Forbes \etal (1996) to examine the
luminosity function of $\sim$350 globular clusters in the central regions of
the giant (M$_V$ = --22.1) elliptical NGC 4365.  In particular, we fit the
globular cluster luminosity function (GCLF) by both a Gaussian and $t_5$
distribution, using the maximum likelihood analysis of Secker \& Harris (1993).
The GCLF is well fit by a turnover magnitude of $m_V^0$ = 24.2 $\pm$ 0.3 and
dispersion $\sigma$ = 1.28 $\pm$ 0.15 (the two fitting profiles give similar
results).  Our results are compared to previous work on the GCLF of NGC 4365
and other Virgo ellipticals. Using the most recent determination of the Milky
Way and M31 galaxy's GCLF turnover magnitude of Sandage \& Tammann (1995), and
a metallicity correction based on the precepts of Ashman \etal (1995), we
derive a distance modulus of 31.6 $\pm$ 0.3. This is in reasonable agreement
with $(m - M)$ = 31.74 $\pm$ 0.16 derived from surface brightness fluctuation
measurements of NGC 4365 and provides further support to the hypothesis that
the absolute turnover magnitude of GCLFs is approximately constant or
`universal' for all galaxies.  Our distance modulus also supports the previous
findings from ground--based data that the GCLF turnover is $\sim$0.7 magnitudes
fainter, or $\sim$6 Mpc more distant, than that of ellipticals in the Virgo
cluster core. As such NGC 4365 may lie in the W$^{'}$ group of the SW extension
of the Virgo cluster or in the background W cloud (Binggeli, Tammann, \&
Sandage 1987).  

Adopting a velocity, with respect to the cosmic microwave background, for NGC
4365 of V$_{CMB}$ = 1592 $\pm$ 24 km s$^{-1}$ gives a Hubble constant of
H$_{\circ}$ = 72 (+10, --12) km s$^{-1}$ Mpc$^{-1}$ (internal errors only).
This value lies between recent determinations of H$_{\circ}$ from the GCLF of
NGC 4881 and M87 using WFPC2.  After correcting for undetected objects, we have
estimated the total number of globular clusters in a 5~h$^{-1}$ kpc radius
circle about the galaxy center to be 911. Using the integrated galaxy light
within this region, we derive a `local' specific frequency of $S$ = 6.4 $\pm$
1.5.  This measure has the advantage of requiring fewer uncertain corrections
than a total estimate. For a distance modulus of 31.74, the ground--based total
specific frequency becomes 5 $\pm$ 2, which is similar to our local $S$ value
and in excellent agreement with the average $S$ value for a large sample of
early type galaxies.  

\noindent
{\bf Acknowledgments}\\
We are particularly grateful to J. Secker for the use of his maximum likelihood
code and useful suggestions. We also thank R. Elson, C. Grillmair and R.
Guzm\'an for helpful discussions. This research was funded by the HST grant
AR-05794.01-94A\\

\newpage
\noindent{\bf References}

\noindent
Ajhar, E. A., Blakeslee, J. P., \& Tonry, J. L. 1994, AJ, 108, 2087 (ABT94)\\
Ashman, K. M., Conti, A., \& Zepf, S. E. 1995, AJ, 110, 1164\\ 
Baum, W. A., \etal 1995, AJ, 110, 2537\\
Binggeli, B, Tammann, G. A., \& Sandage, A. 1987, AJ, 94, 251\\
Brodie, J. P., \& Huchra, J. 1991, ApJ, 379, 157\\
Burrows, C., \etal 1993, Hubble Space Telescope Wide Field and
Planetary Camera 2 Instrument Handbook, STScI\\
Carollo, C. M., Franx, M., Illingworth, G. D., \& Forbes, D. A. 1996,
ApJ, submitted\\
Couture, J., Harris, W. E., \& Allwright, J. W. B., 1990, ApJS, 73,
671\\
Davies, R. L., \etal 1987, ApJS, 64, 581\\
Faber, S. M., \etal 1989, ApJS, 69, 763\\
Forbes, D. A. 1994, AJ, 107, 2017\\
Forbes, D. A., Elson, R. A. W., Phillips, A. C., 
Illingworth, G. D. \& Koo, D. C. 1994, ApJ, 437, L17\\
Forbes, D. A., Franx, M., Illingworth, G. D., \& Carollo, C. M. 1996,
ApJ, in press\\
Fleming, D. E. B., Harris, W. E., Pritchet, C. J., \& Hanes,
D. A. 1995, AJ, 109, 1044\\
Goudfrooij, P., Hansen, L., Jorgensen, H. E., \& Norgaard Nielson, H. U.  
1994, A \& AS, 105, 341\\
Harris, W. E. 1991, ARAA, 29, 543\\
Harris, W. E., Allwright, J. W. B., Pritchet, C. J., \& 
van den Bergh, S. 1991, ApJS, 276, 491\\
Huchra, J., Davis, M., Latham, D., \& Tonry, J. 1983, ApJS, 52, 89\\
Jacoby, G. H., \etal 1992, PASP, 104, 599 (J92)\\
Perelmuter, J. L. 1995, ApJ, 454, 762\\
Sandage, A., \& Tammann, G. A. 1995, ApJ, 446, 1\\
Secker, J. 1992, AJ, 104, 1472\\
Secker, J., \& Harris, W. E. 1993, AJ, 105, 1358 (SH93)\\
Stetson, P. B., 1987, PASP, 99, 191\\
Surma, P. 1992, Structure, Dynamics and Chemical Evolution of
Elliptical Galaxies, ed. I. J. Danziger, W. W. Zeilinger and K. Kjar,
ESO: Garching, p. 669\\
Tonry, J. L., Ajhar, E. A., \& Luppino, G. A. 1990, AJ, 100, 1416\\
Tonry, J. L. 1991, ApJ, 373, L1\\
van den Bergh, S. 1995, AJ, 110, 2700\\
Whitmore, B. C., Sparks, W. B., Lucas, R. A., Macchetto, F. D., \&
Biretta, J. A. 1995, ApJ, in press (W95)\\


\begin{figure*}[p]
\centerline{\psfig{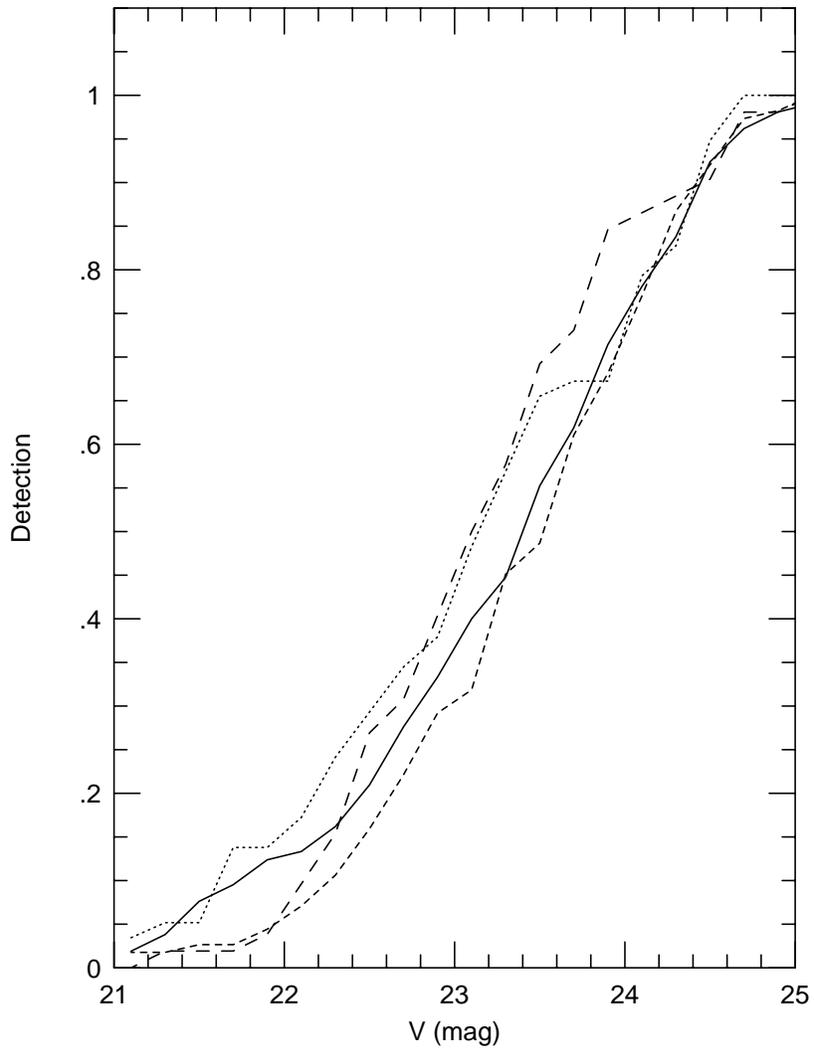}}
\caption{\label{fig1}
Detection of actual globular clusters in each CCD normalized at V = 25. The
different CCDs are represented as follows: PC (long dashed), WF2 (short dash),
WF3 (dotted) and WF4 (solid). The fraction of actual detected globular clusters
in all 4 CCDs is similar to V = 25.
}
\end{figure*}

\begin{figure*}[p]
\centerline{\psfig{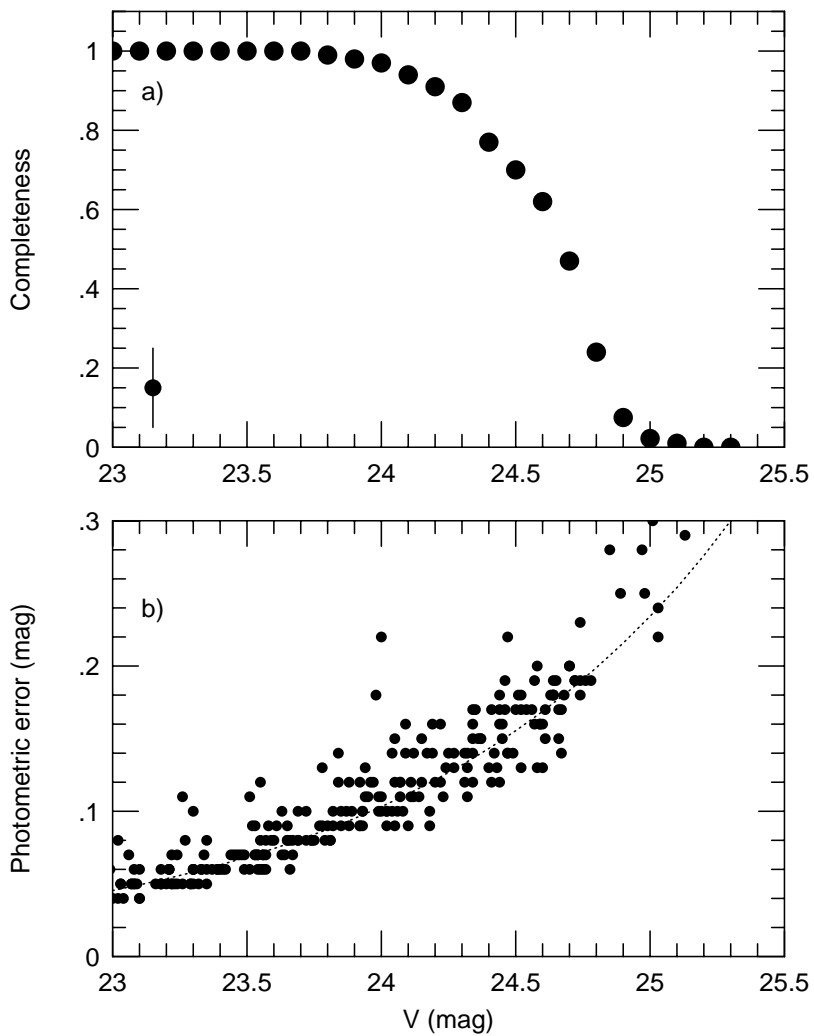}}
\caption{\label{fig2}
Completeness function for GC detection from simulations. Circles show the
fraction of simulated GCs detected in 0.1 magnitude bins. A typical error bar
is shown in the lower left.  {\bf b)} Photometric error as a function of GC V
magnitude determined from DAOPHOT. Circles show the data points, and the dashed
line an exponential fit to the data of the form p.e. = exp~[a~(V~--~b)].
}
\end{figure*}

\begin{figure*}[p]
\centerline{\psfig{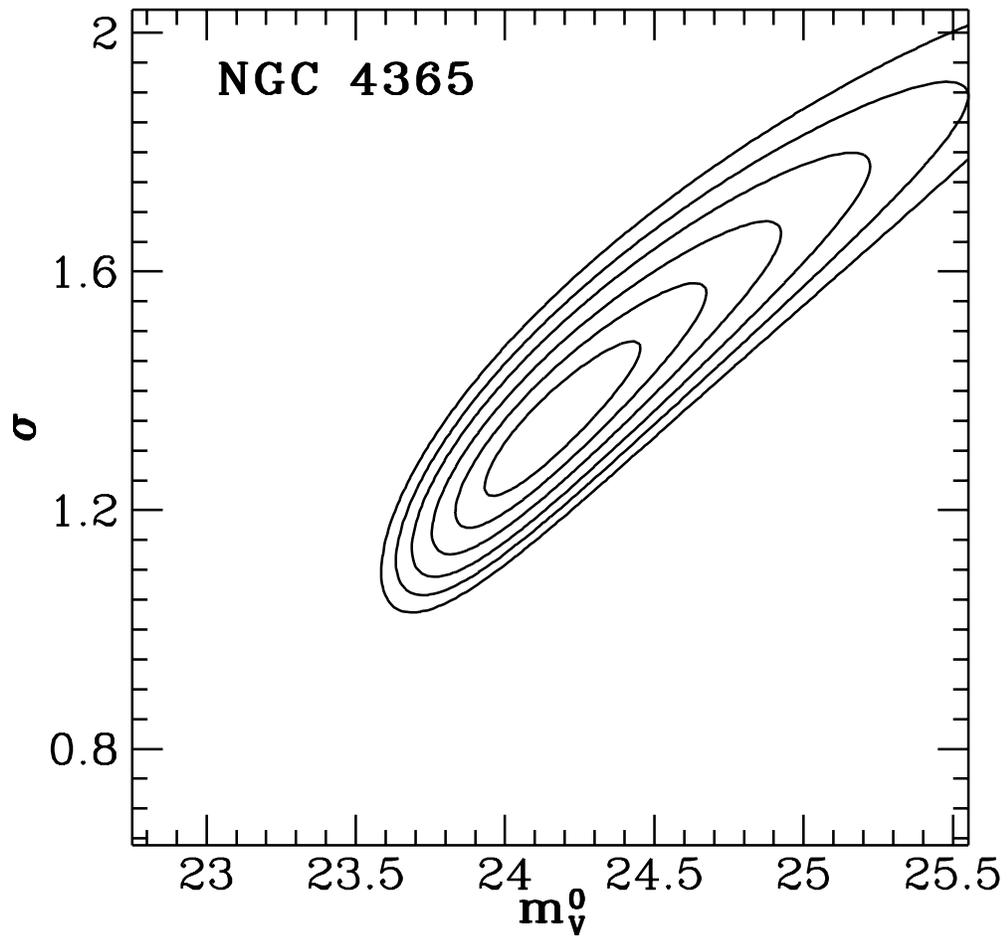}}
\caption{\label{fig3}
Probability contours for the turnover magnitude and dispersion for a Gaussian
fit from the maximum likelihood code of Secker \& Harris (1993). Contours
represent 0.5 to 3 standard deviations probability limits from the best
estimate (see Table 1).
}
\end{figure*}

\begin{figure*}[p]
\centerline{\psfig{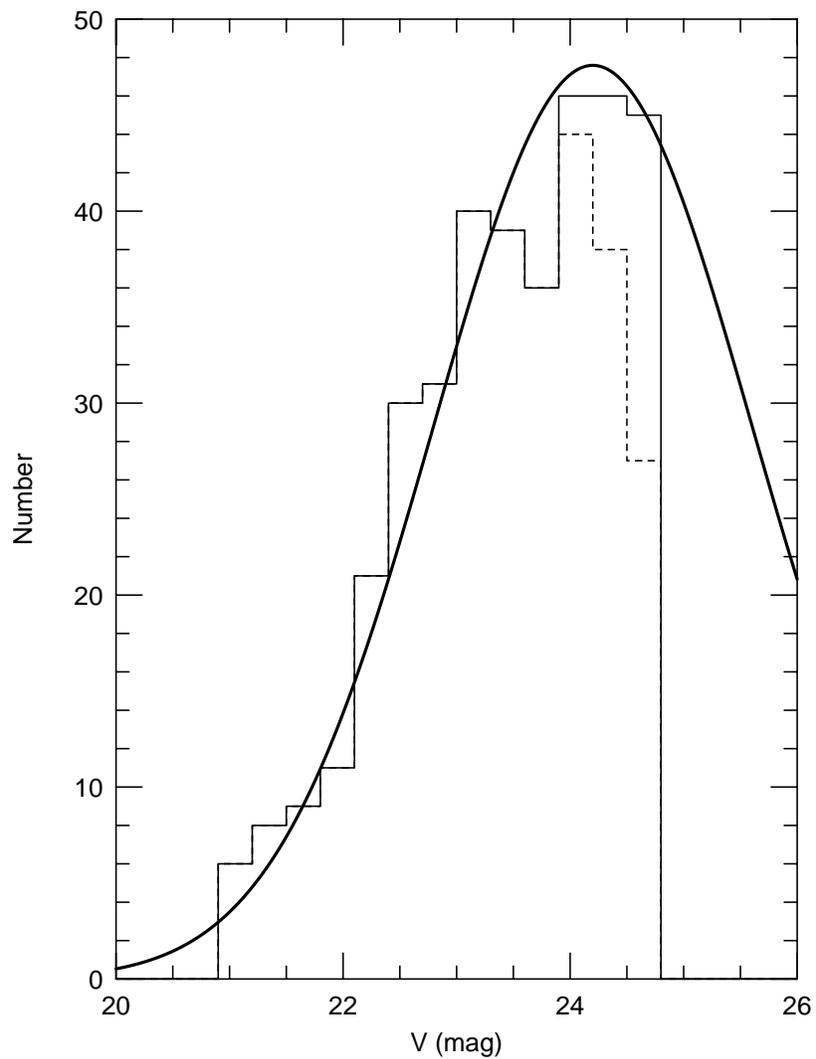}}
\caption{\label{fig4}
Globular cluster luminosity function for NGC 4365. The raw data is shown by a
dashed line, and by a thin solid line after completeness correction has been
applied. The maximum likelihood best fit Gaussian profile, which includes the
effects of photometric error and background contamination, is superposed as a
thick solid line.  Note that the fitting procedure does not use binned data.  
}
\end{figure*}

\clearpage
\begin{figure*}
\centerline{\psfig{figure=table1.epsi,width=300pt}}
\end{figure*}

\begin{figure*}
\centerline{\psfig{figure=table2.epsi,width=300pt}}
\end{figure*}

\end{document}